\documentclass[a4paper]{PoS}
\pdfoutput=1

\usepackage{graphicx}
\usepackage{amsmath}
\usepackage{amssymb}

\newcommand{\la}{\left\langle}
\newcommand{\ra}{\right\rangle}
\newcommand{\hk}{\text{HK}}
\newcommand{\SU}{\text{SU}}
\DeclareMathOperator{\Tr}{Tr}
\newcommand{\as}{\text{asym}}

\title{Numerical study of large-N phase transition of smeared Wilson loops in 4D pure YM theory}

\ShortTitle{Large-$N$ phase transition of smeared Wilson loops in 4D pure YM}

\author{\speaker{Robert Lohmayer}%
        \\
        Rutgers University, Department of Physics and Astronomy, Piscataway,
        NJ 08854, USA \\
        E-mail: \email{lohmayer@physics.rutgers.edu}}

\author{Herbert Neuberger\\
        Rutgers University, Department of Physics and Astronomy, Piscataway,
        NJ 08854, USA\\
        E-mail: \email{neuberg@physics.rutgers.edu}}

\abstract{
In Euclidean four-dimensional $\SU(N)$ pure gauge theory, eigenvalue
distributions of Wilson loop parallel transport matrices around closed
spacetime curves show non-analytic behavior (a 'large-$N$ phase transition')
at a critical size of the curve. We focus mainly on an observable composed of traces of the Wilson loop operator in all totally antisymmetric representations, which is regularized with the help of smearing. By studying sequences of square Wilson loops on a hypercubic lattice with standard Wilson action, it is shown that this observable has a nontrivial continuum limit as a function of the physical size of the loop. We furthermore present (preliminary) numerical results confirming that, for large $N$, the $N$ dependence in the critical regime is governed by the universal exponents $1/2$ and $3/4$ as expected (Burgers universality).
}

\FullConference{XXIX International Symposium on Lattice Field Theory \\
         July 10-16,  2011\\
         Squaw Valley, Lake Tahoe, California}

\begin{document}

\section{Introduction}

Already in 1981, Durhuus and Olesen \cite{Durhuus:1980nb} discovered that Wilson loops in pure
$\SU(N)$ gauge theory in two Euclidean dimensions exhibit a transition from an
'ordered phase' to a 'disordered phase' at large $N$. In two dimensions, the
eigenvalue spectrum of the untraced Wilson loop unitary matrix depends only on
the area that is enclosed by the (nonselfintersecting) space-time curve defining the loop and a
sharp infinite-$N$ phase transition occurs at a critical value of this
area. At this point, the gap in the spectrum, that is present for small loops,
closes in a non-analytic way.

In Ref.~\cite{Narayanan:2006rf}, Narayanan and Neuberger conjectured  that a similar large-$N$
transition also occurs in three and four dimensions and that all transitions
belong to a single universality class. If this is indeed the case, it might
be possible to connect the perturbative regime with
non-perturbative models in four dimensions using the universal properties of
the transition at a critical scale. However, so far the universality
conjecture has been confirmed only for the three-dimensional case \cite{Narayanan:2007dv}. 

In the following, we present a
numerical study of the transition and its universal properties in four dimensions using lattice
methods. 

\section{Main observable}

The Wilson loop matrix associated with a closed spacetime curve $\mathcal C$
is defined as usual:
\begin{align}
W(\mathcal C)=\mathcal{P} \exp\left(i\oint_{\mathcal{C}} A_\mu(x) dx_\mu
\right) \in \SU(N)
\end{align}
with path-ordering operator $\mathcal{P}$. 

To study the large-$N$ transition in the spectrum of $W$, we focus on 
\begin{align}
\mathcal{O}_N(y,\mathcal C)&=\la \det\left(e^{\frac y2}+e^{-\frac y 2 }
W_f(\mathcal C)\right)
\ra = \sum_{k=0}^N e^{\left(\frac N2 -k\right)y}\la\chi_k^\as(W(\mathcal C))\ra\,,
\end{align}
where all totally antisymmetric representations of $W$ enter ($f$ denotes the
fundamental representation and $\chi_k^\as$
denotes the character in the $k$-fold totally antisymmetric representation of
$\SU(N)$). The variable $y$ is defined such that a zero of the determinant at $y=0$ corresponds
to an eigenvalue-angle $\theta=\pm \pi$ on the complex unit circle. The gap in
the eigenvalue spectrum closes at $\theta=\pm \pi$
(cf.~Fig.~\ref{fig-ShockRhoInfN} below). Therefore, useful information about the large-$N$ non-analyticity can be obtained
by expanding $\mathcal{O}_N(y,\mathcal C)$ in powers of $y$ around $y=0$ (with
$\theta_{CP}=0$, only even powers of $y$ enter):
\begin{align}
\mathcal{O}_N(y, \mathcal C)= a_0(\mathcal C) + a_1(\mathcal C) y^2 + a_2(\mathcal C) y^4 +O\left(y^6\right). 
\end{align}
In particular, it turns out (see below) that the scaling-invariant ratio
\begin{align}
\omega(\mathcal C)=\frac{a_0(\mathcal C) a_2(\mathcal C)}{a_1(\mathcal C)^2}
\end{align}
provides a valuable signal for the large-$N$ phase transition in the spectrum
of $W$.

\section{Universality class from heat-kernel model}
The heat-kernel probability density (w.r.t.~the Haar measure) for an $\SU(N)$
matrix $W$ is given by
\begin{align}\label{eq-HKprob}
\mathcal{P}_N^\hk(W,t)&=\sum_{\text{all irred.}\ r} d_r \chi_r(W) e^{-\frac tN C_2(r)}\,,
\end{align}
implying
\begin{align}
\la \chi_r(W) \ra_\hk = d_r e^{-\frac tN C_2(r)}
\end{align}
due to character orthogonality. The parameter $t$ can be interpreted as
a diffusion time, $d_r$ and $C_2(r)$ denote the dimension and the quadratic
Casimir invariant of the irreducible representation $r$. 
The above probability density can be realized in a simple multiplicative
random matrix model and holds exactly in two-dimensional pure Yang-Mills
theory when $t$ is identified with the dimensionless area variable $g^2 N
\mathcal A$ (with 't~Hooft coupling $g^2N$ and $\mathcal A$ denoting the area
enclosed by the loop). 
The heat-kernel single eigenvalue-angle distribution is
given by \cite{Lohmayer:2009aw}
\begin{align}
\rho_N^{\hk}(\theta, t)=\frac{1}{2\pi}+\frac{1}{\pi N}\sum_{p=0}^{N-1} (-1)^p
\sum_{q=0}^\infty d(p,q) e^{-\frac{t}{N} C(p,q)}\cos((p+q+1)\theta)
\end{align}
with $C(p,q)=\frac{1}{2}(p+q+1)\left ( N -\frac{p+q+1}{N} + q-p\right )$ and 
$d(p,q)=\frac{(N+q)!}{p!q!(N-p-1)!}\frac{1}{p+q+1}$.

\begin{figure}[htb]
    \includegraphics[width=0.4\textwidth]{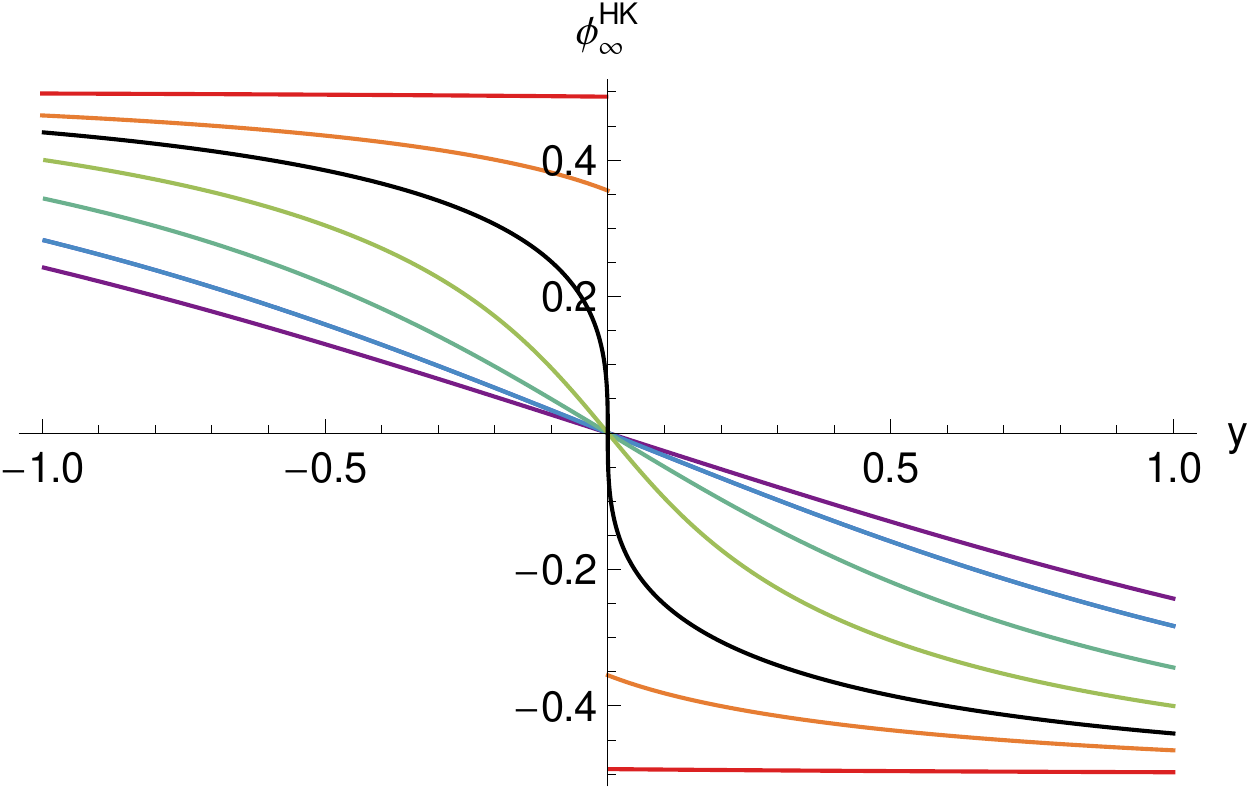}\hfill
    \includegraphics[width=0.1\textwidth]{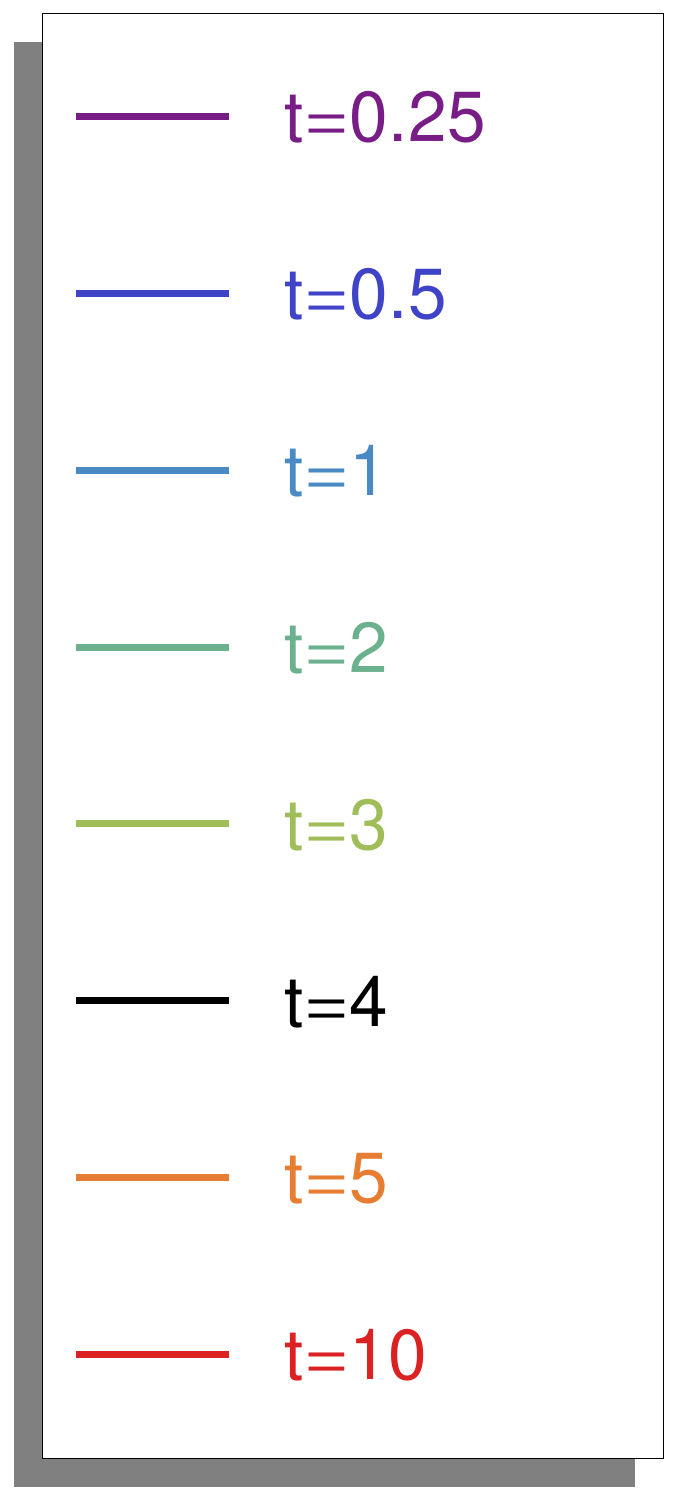}\hfill
    \includegraphics[width=0.4\textwidth]{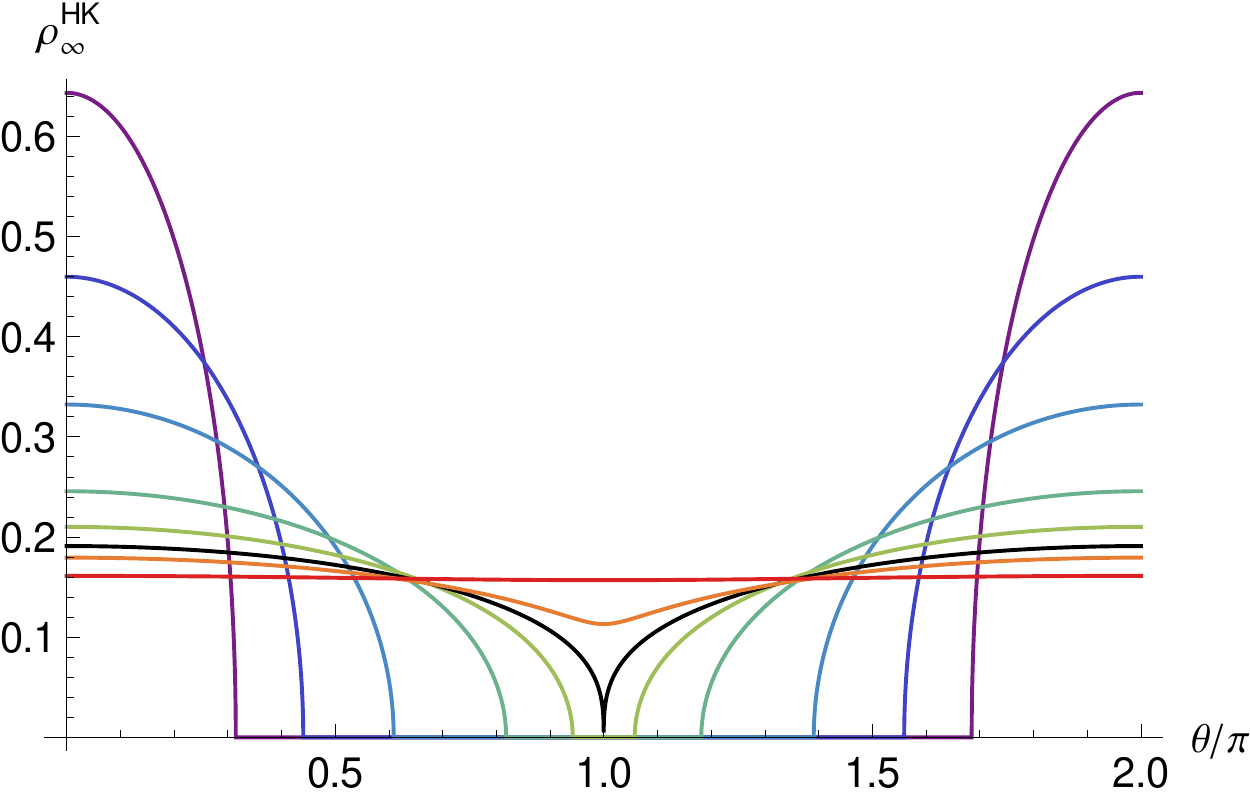}
\caption{$\phi_N^\hk$ (left) and heat-kernel single eigenvalue density
  $\rho_N^\hk$ (right) in the limit $N\to\infty$ for various $t$.}
\label{fig-ShockRhoInfN}
\end{figure}

An immediate consequence of the group theoretical structure \eqref{eq-HKprob} is that  
\begin{align}
\phi_N^\hk(y,\tau)= -\frac 1 N \frac\partial{\partial y} \log \mathcal
O_N^\hk(y,\tau)
= -\frac 1 N \frac\partial{\partial y} \log \la \det\left(e^{\frac y2}+e^{-\frac y 2 }
W_f\right)
\ra_\hk
\end{align}
with $\tau=t(1+1/N)$ satisfies Burgers' equation \cite{Neuberger:2008mk} 
\begin{align*}
\partial_\tau \phi_N^\hk+\phi_N^\hk\partial_y\phi_N^\hk=\frac1{2N}\partial_y^2 \phi_N^\hk
\end{align*}
with initial condition $\phi_N^\hk(y,0)=-\frac 12 \tanh \frac y2$.
At $N=\infty$, Burgers' equation produces a 'shock-wave' singularity at $y=0$
when $\tau$ reaches the critical value $\tau_c=4$, cf.~Fig.~\ref{fig-ShockRhoInfN}.

From Burgers' equation (at finite $N$), we get
\begin{align}
\partial_\tau \frac1{\partial_y \phi_N^\hk |_{y=0}}=3\omega_N^\hk(\tau)-\frac 12\,.
\end{align}
At infinite $N$, the inverse slope of $\phi_\infty^\hk$ at $y=0$ increases from $-4$ at $\tau=0$ to $0$ at
  $\tau=4$ since
\begin{align*}
\omega_\infty^\hk(\tau)=\begin{cases} 1/2,\qquad 0\leq\tau<4, \\ 1/6, \qquad  \tau>4, \end{cases}
\end{align*}
resulting in a discontinuous jump in $\phi_\infty^\hk(y)$ for $\tau>4$ at $y=0$. 
This discontinuity results in a nonzero single
eigenvalue density at $\theta=\pm \pi$ (on the unit circle in the complex
plane). The
  infinite-$N$ phase transition (non-analytic behavior with characteristic
  exponents) in the eigenvalue density of $W$ occurs at the point where the gap in
  the spectrum disappears (this happens at $\tau=\tau_c=4$), cf.~Fig.~\ref{fig-ShockRhoInfN}.

\begin{figure}[htb]
\begin{center}
    \includegraphics[width=0.5\textwidth]{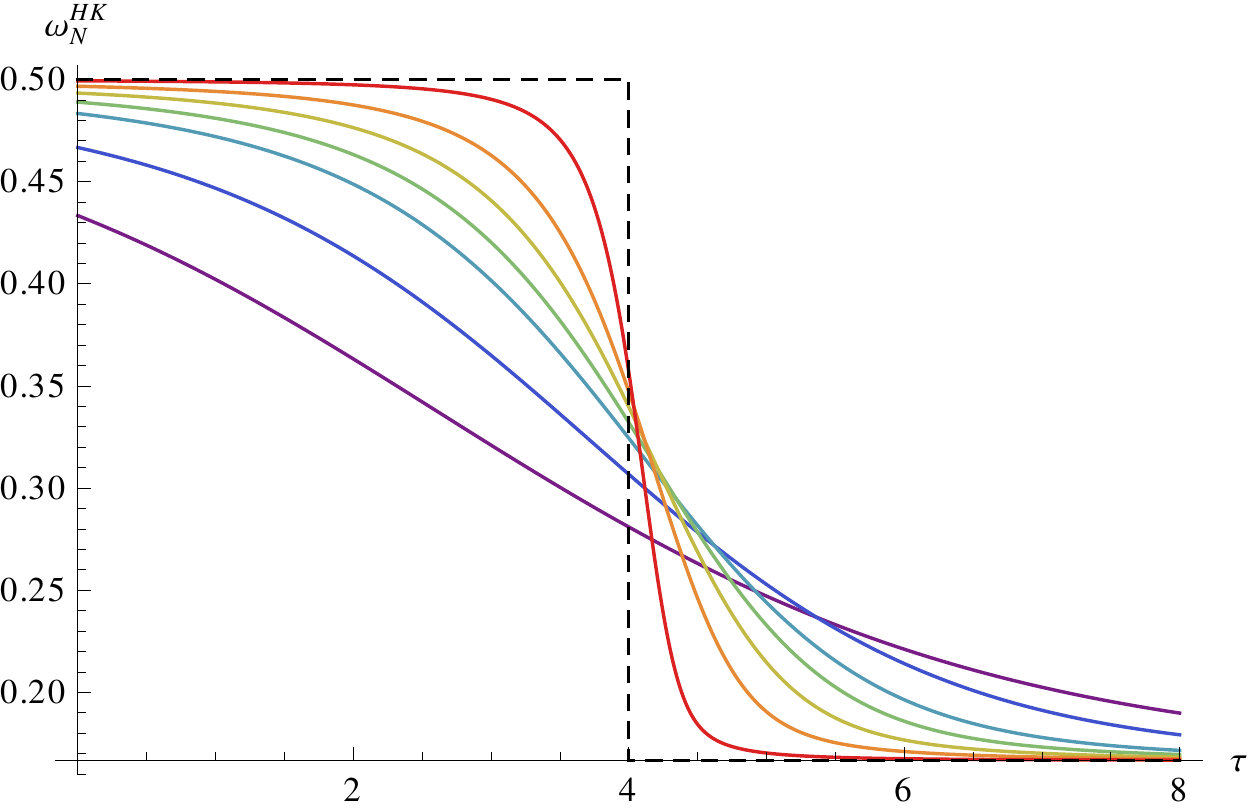}
\qquad \includegraphics[width=0.12\textwidth]{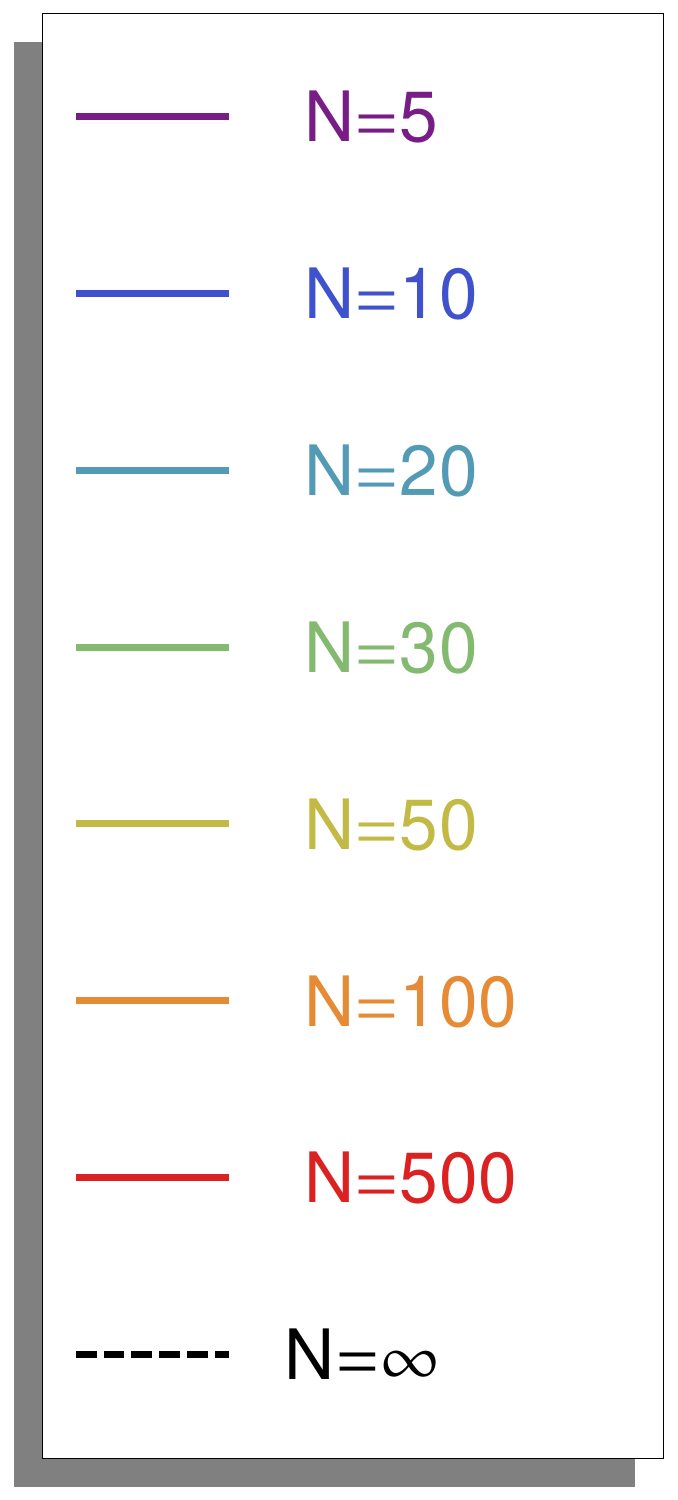}
\caption{$\omega_N^\hk(\tau)$ for various $N$.}    
\label{fig-omegatau}
\end{center}
 \end{figure}

The singularity is absent for any finite $N$, and very large values of $N$ are
needed to see its formation, cf.~Fig.~\ref{fig-omegatau}.
From Burgers' equation we obtain:
\begin{align}
\lim_{N\to\infty} N^{-\frac 32} \left. \frac{a_1}{a_0}\right|_{\tau=4}&=\frac 18 \sqrt{\frac
  32} \frac 1K, \qquad\qquad K \equiv\frac{1}{4\pi} \Gamma^2\left( \frac{1}{4}
\right ) \approx 1.046,\\
\lim_{N\to\infty} N^{-\frac32} \left.\frac{a_2}{a_1}\right |_{\tau=4} & = \frac{1}{24}\sqrt{\frac{3}{2}}K,\\
\lim_{N\to\infty}  \left. \omega_N \right |_{\tau=4} & =\frac{1}{3}
K^2,\\
\lim_{N\to\infty} N^{-\frac12} \left. \frac{d\omega_N}{d\tau} 
\right |_{\tau=4}  &=-\frac{1}{6} \sqrt{\frac{3}{2}} K(K^2-1).
\end{align}
Furthermore, the (purely imaginary) roots of $\mathcal O_N^\hk(y,\tau)$ in the critical regime (around $y=0$,
$\tau=4$) scale like $N^{-\frac 34}$.

\section{Numerical results in 4D}

In four dimensions, the Wilson loop operator develops a perimeter divergence
(and additional corner divergences if the spacetime curve has kinks). A
convenient way of regularization is by smearing (cf.~Ref.~\cite{Lohmayer2011a} for details). We deal with the
intrinsic UV divergences of the action by using a lattice discretization of
the theory. To extrapolate to the continuum, we compute sequences of square
Wilson loops of sides $1\leq L \leq 9$ for inverse 't Hooft couplings
  $0.348\leq b=\frac1{g^2N} \leq 0.374$ at $N=19$, using a single plaquette Wilson action on a hypercubic lattice (volumes
  $12^4$ and $14^4$ are used to exclude data contaminated by finite-volume effects) and a combination of heat-bath and overrelaxation
updates. The amount of smearing $S$ is chosen proportional to the size of the
loop, $S=L^2/55$. We then measure $\omega_N(b,L)$ (determined from the average characteristic polynomial) as a signal for the large-$N$ phase transition.
Large-$N$ scaling is tested with additional runs at $N=11$.

\begin{figure}
\begin{center}
\includegraphics[width=0.5\textwidth]{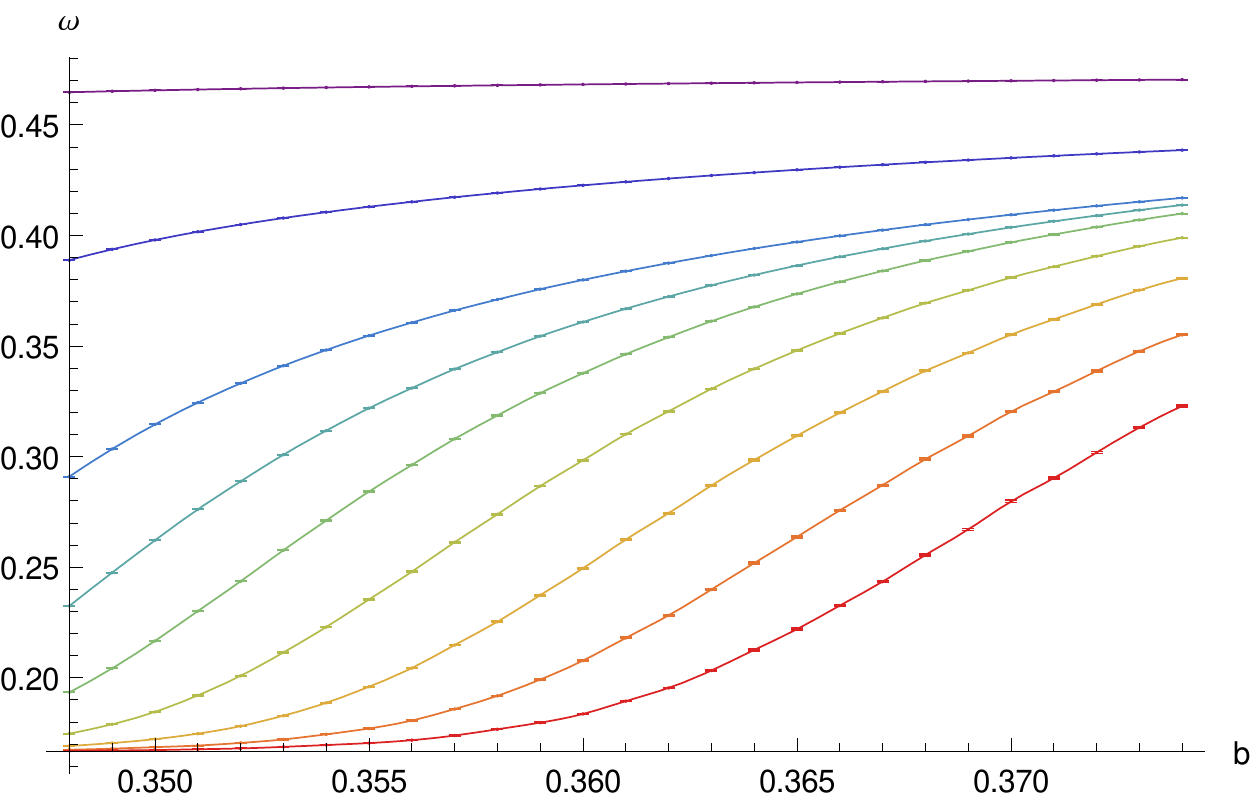}
\qquad \includegraphics[width=0.1\textwidth]{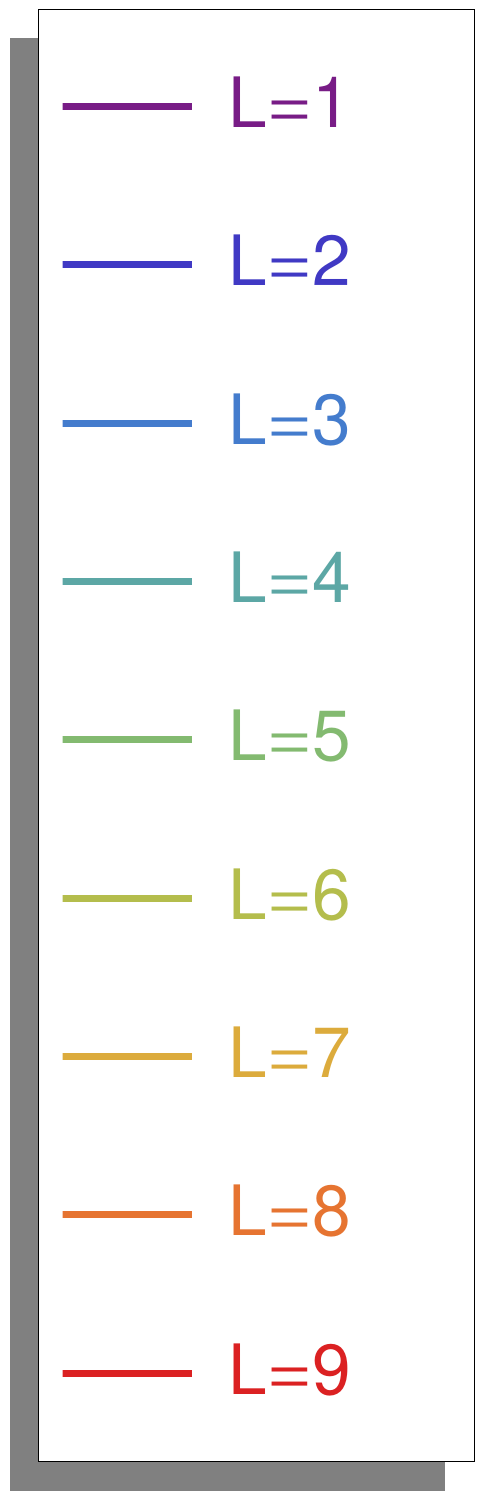}
\caption{Lattice results for $\omega_{19}(b,L)$ (with cubic spline
  interpolation between data points).}
\label{fig-Omega}
\end{center}
\end{figure}

\begin{figure}
\begin{center}
\includegraphics[width=0.44\textwidth]{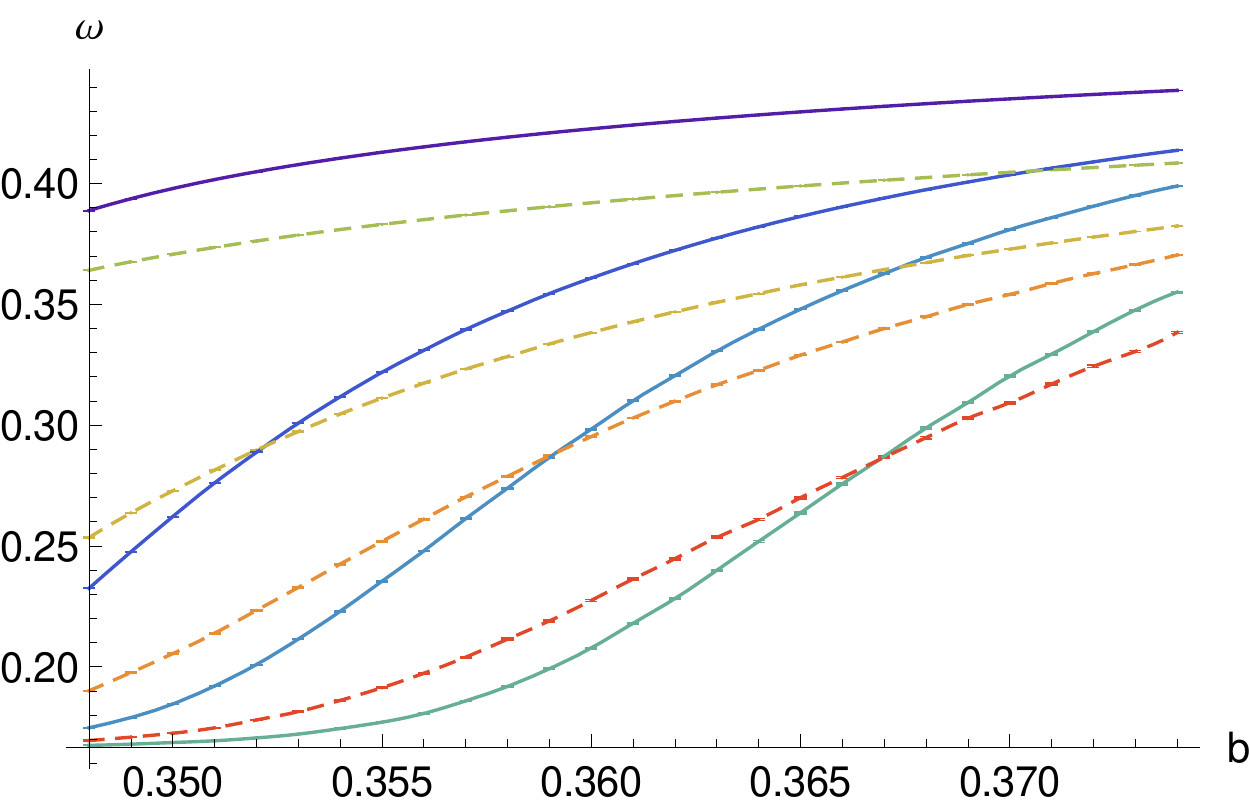}\hfill
\includegraphics[width=0.1\textwidth]{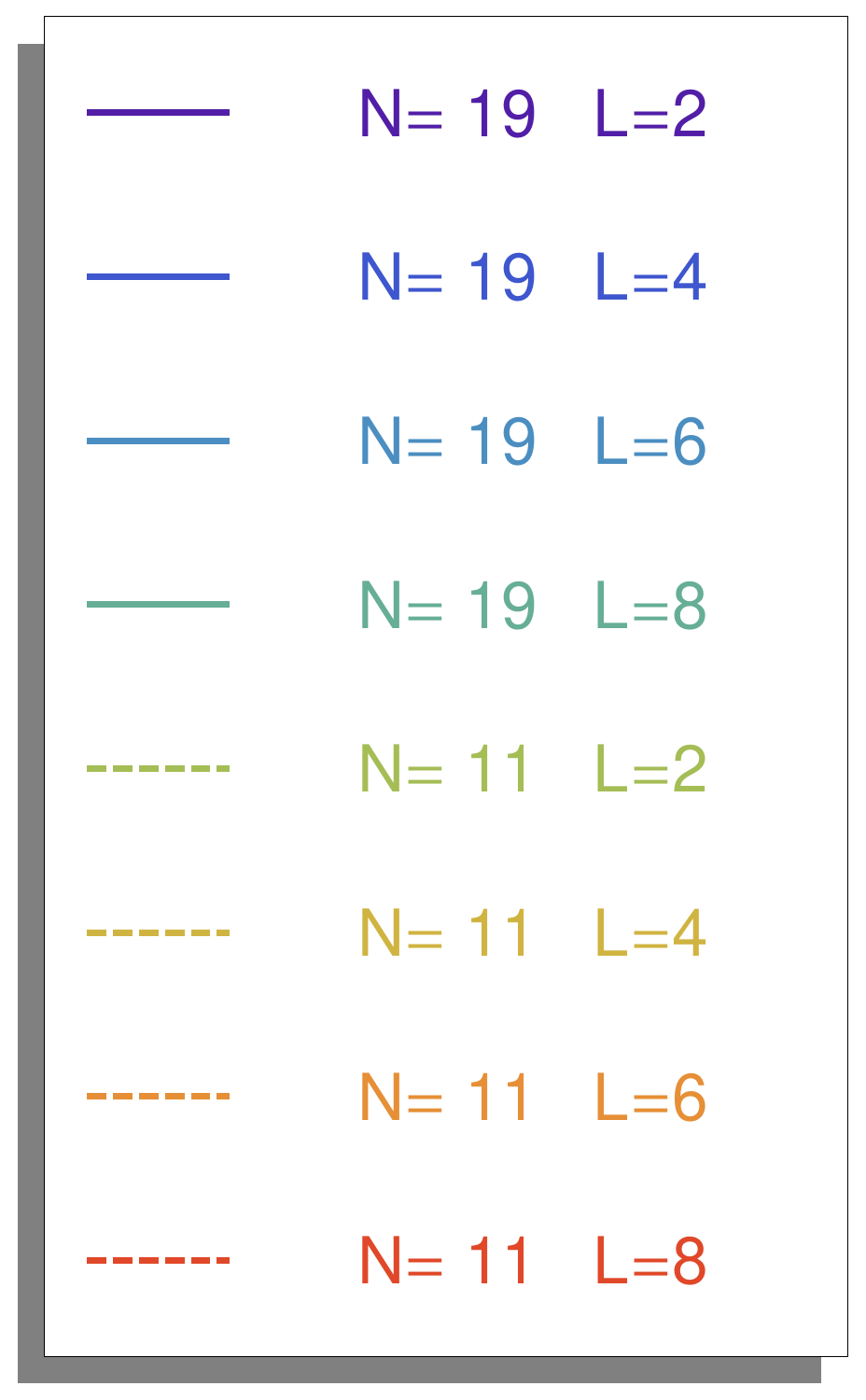}\hfill
\includegraphics[width=0.44\textwidth]{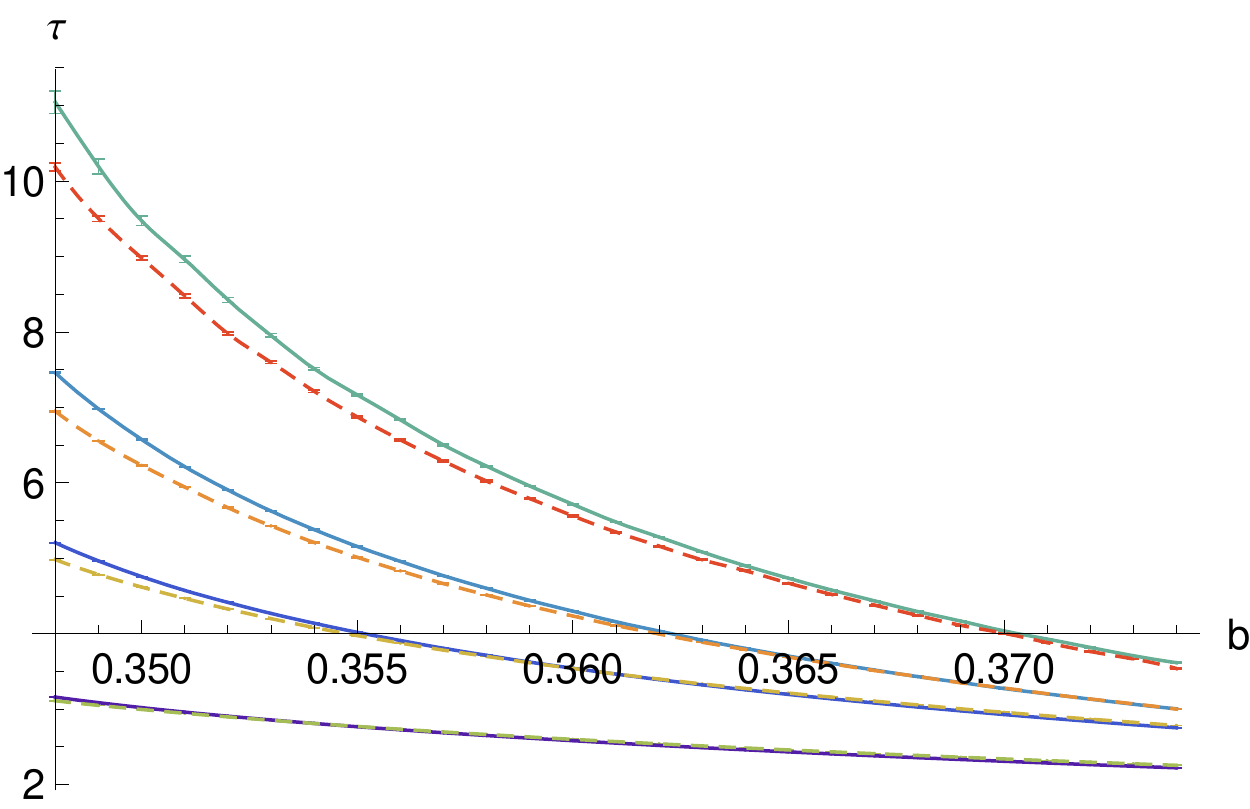}
\caption{$\omega_N(b,L)$ and $\tau_N(b,L)$ for $N=19$ (solid) and $N=11$
  (dashed) and $L=2,4,6,8$.}
\label{fig-OmegabTaub}
\end{center}
\end{figure}

Figure~\ref{fig-Omega} shows our numerical results for
$\omega_{19}(b,L)$. Similar to the heat-kernel model, we observe that very
large values of $N$ would be needed to allow for a direct observation of  the singular large-$N$ behavior in
$\omega_N$. Therefore, we define an $N$-dependent map $\omega_N(b,L) \to \tau_N(b,L)$ through
\begin{align}
\omega_N(b,L)=\omega_{N}^\hk(\tau_N(b,L))\,.
\end{align}     
The required inversion is unique and the map can be used even though the
$r$-dependence of $\la \chi_r(W) \ra$ in 4D
differs from exact Casimir scaling. While the formation of the jump in
$\omega_N$ is slow, $\tau_N$ converges rapidly to $\tau_\infty$, cf.~Fig.~\ref{fig-OmegabTaub}. This implies
that, to some subleading order in $\frac 1N$, $\omega_N(b,L)\approx
\omega_N^\hk(\tau_\infty(b,L))$. 

\begin{figure}
\begin{center}
    \includegraphics[trim = 0mm 0mm 0mm 6mm, clip, width=0.5\textwidth]{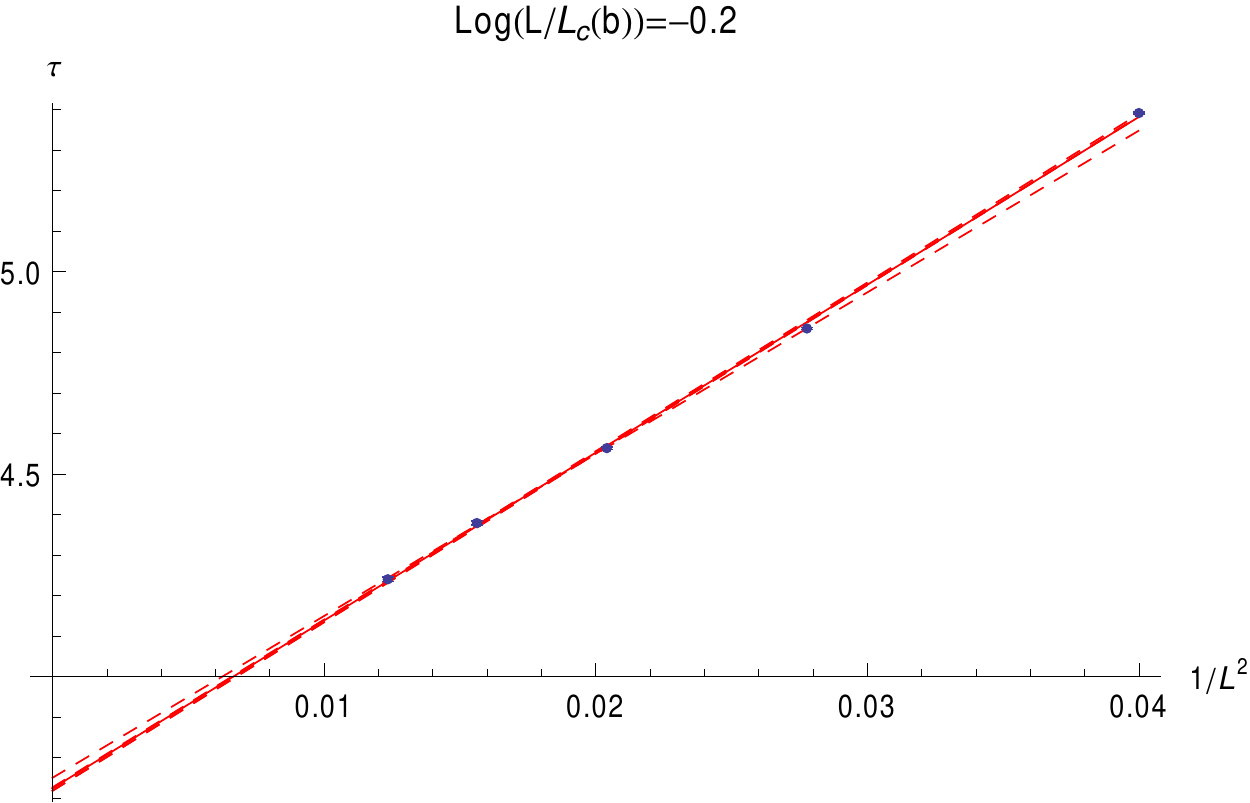}
\caption{Example for continuum extrapolation at $N=19$ and $\log(\frac L{L_c})=-0.2$.} 
\label{fig-ContExtr}
\end{center}
\end{figure}

This relation can be taken over to the continuum limit, where the two
variables $b$ and $L$ get replaced by a single length variable $l=L/L_c(b)$, the side
of the loop in physical units. To set the scale, we use the critical deconfinement
temperature $1/L_c(b)$, determined by
\begin{align}
L_c(b)=0.26 \left(\frac{11}{48 \pi^2 b_i(b)}\right)^{\frac{51}{121}}
e^{\frac{24\pi^2}{11} b_i(b)},\qquad b_i(b)=\frac bN \la \Tr
W_{\text{1x1}} \ra\,.
\end{align}  
The continuum limit of $\tau_N$ is obtained by extrapolating $\tau_N(b,L)$ at
fixed $l$ to $L\to\infty$, cf.~Fig.~\ref{fig-ContExtr}. 

\begin{figure}
\begin{center}
\includegraphics[width=0.5\textwidth]{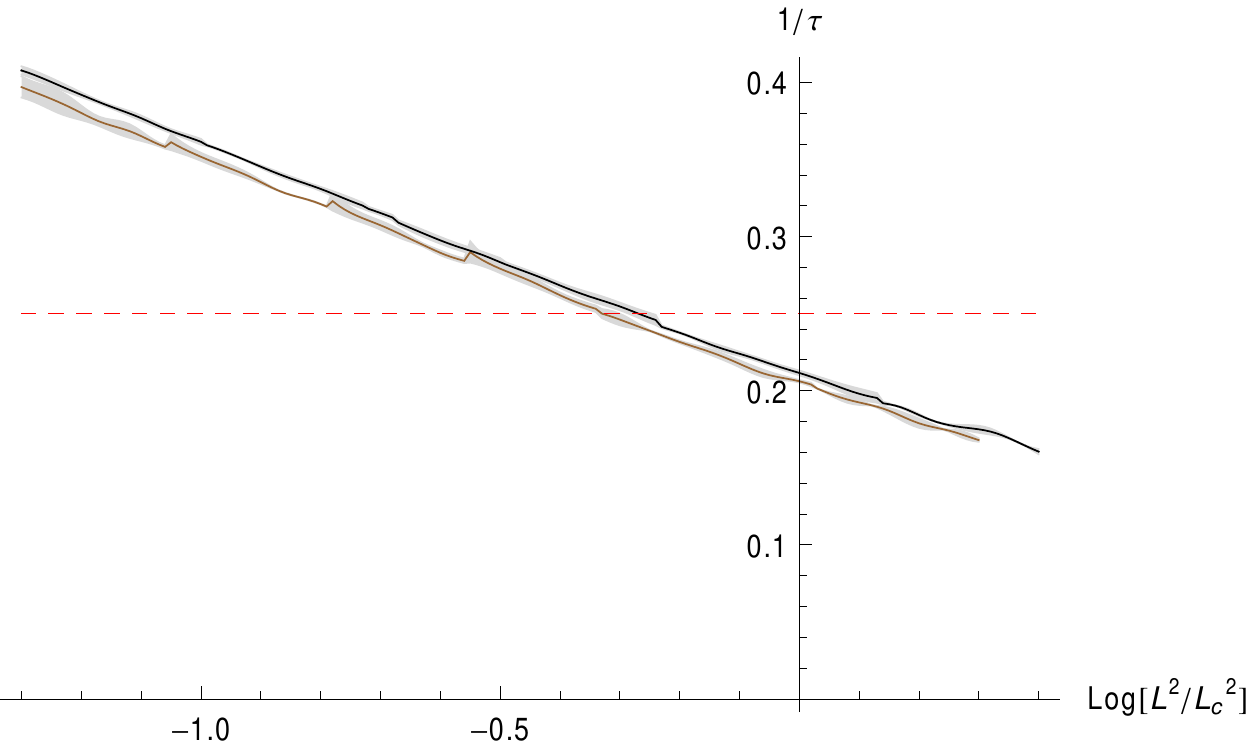}
\end{center}
\caption{Continuum functions $\tau_N(l)^{-1}$ for $N=19$ (black) and $N=11$
  (brown). The gray bands show the corresponding error estimates (jackknife), the horizontal dashed line corresponds to the critical value $\tau_c=4$.}
\label{fig-InvTau}
\end{figure}

We observe that $\tau_N$ has a nontrivial continuum limit which is a smooth
function of the physical loop size $l$ for all $N$, cf.~Fig.~\ref{fig-InvTau}.  
This establishes the transition and its universality since  we can replace
$\mathcal O_N(y,\mathcal C)$ by $\mathcal O_N^\hk(y,\tau_\infty(l))$ in the
vicinity of the critical point ($y=0$, $\tau=4$)  without changing the
singular large-$N$ properties. The dependence on $l$ is consistent with
asymptotic freedom as $\tau$ modulo a shape-dependent factor can be
interpreted as an effective coupling constant.

\begin{center}
\begin{figure}
\includegraphics[width=0.46\textwidth]{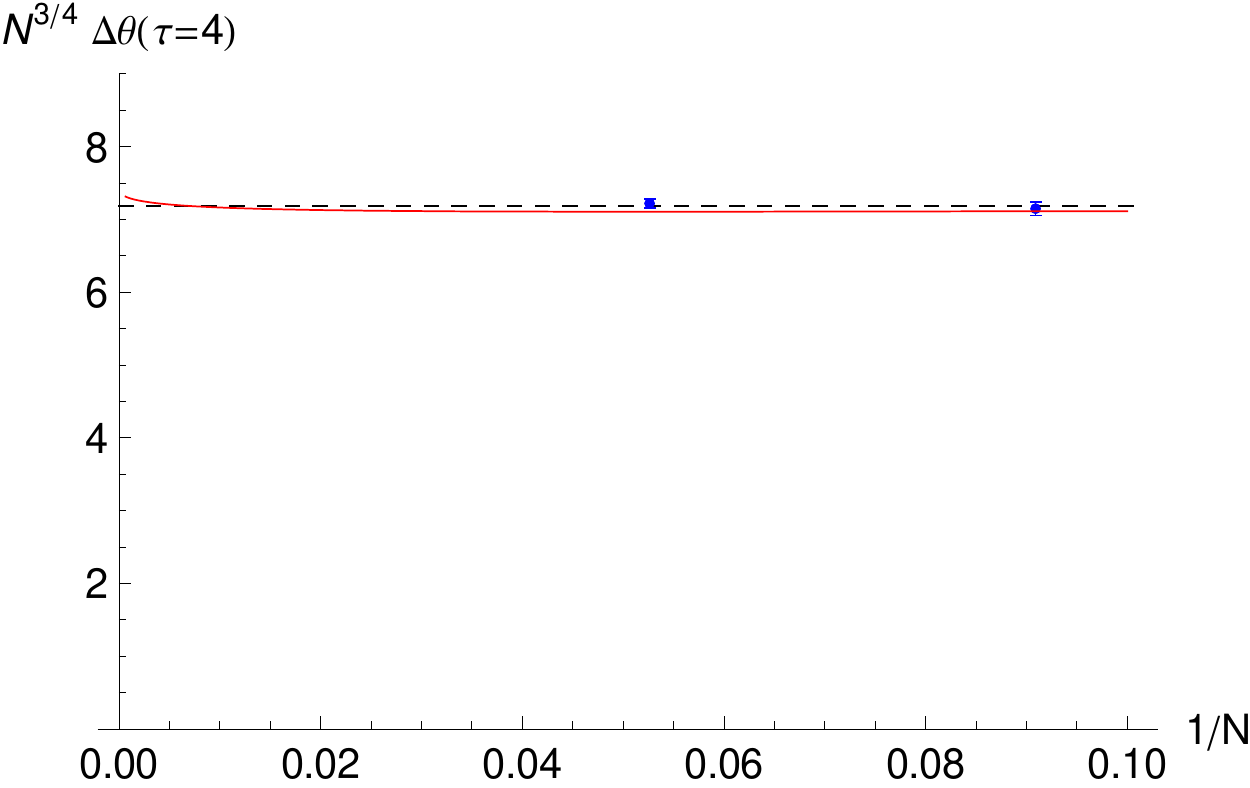}\hfill
\includegraphics[width=0.46\textwidth]{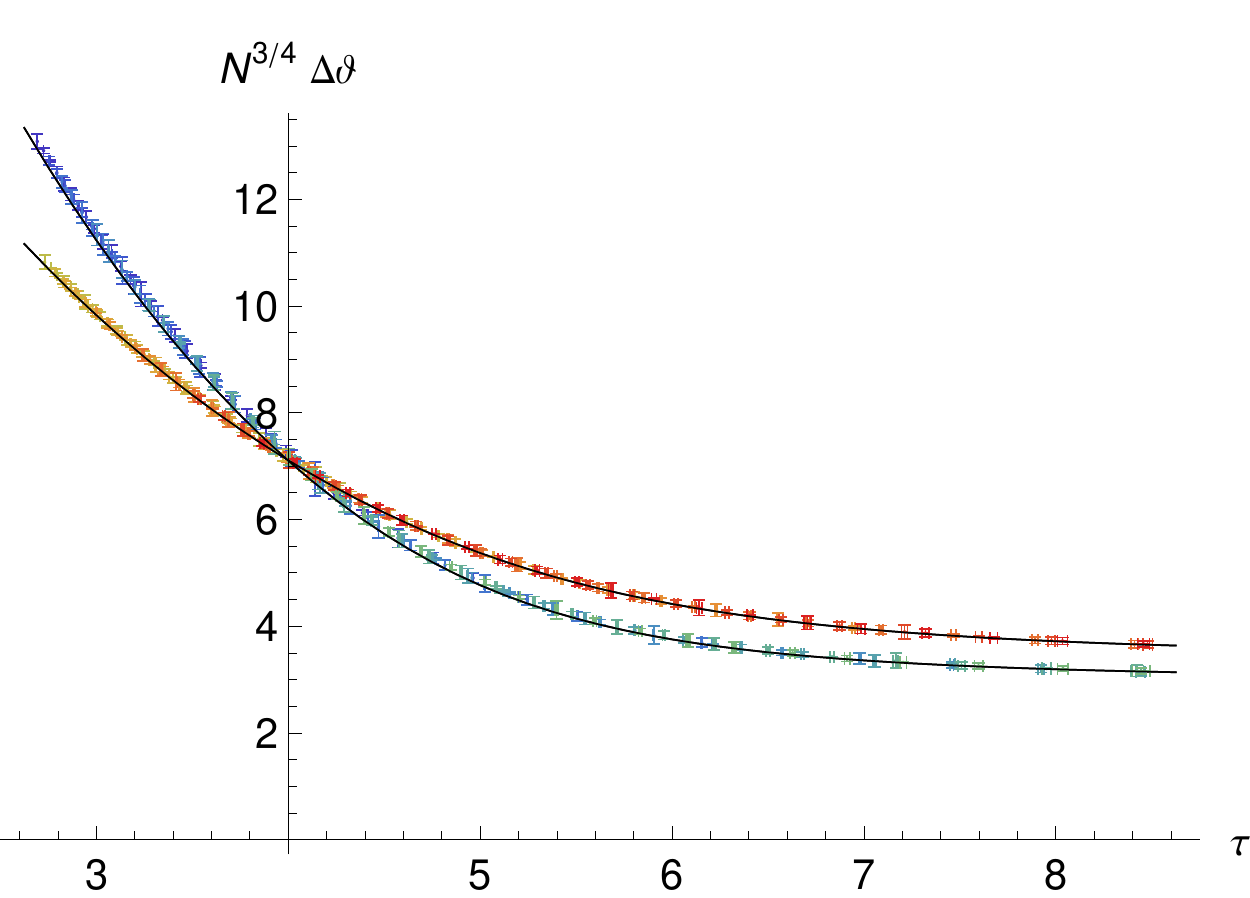}
\caption{Left: appropriately scaled angular difference $\Delta\theta$ between the two peaks in the single
eigenvalue density closest to $\theta=\pi$ at $\tau=4$ as a function of $1/N$
(blue: data points, red: heat-kernel value).\newline
Right: $N^{3/4} \Delta\theta$ as a function of $\tau$ (blue-green points for
$N=19$, red-yellow points for $N=11$; black lines show the corresponding
heat-kernel functions). At the intersection point (at $\tau=4$), $\Delta\theta
\propto N^{-3/4}$.}  
\label{fig-PeakPos}
\end{figure}
\end{center}

Additional checks of universality can be obtained from the critical exponent $3/2$
of the ratio $a_0/a_1$ and the $N^{-3/4}$-scaling of the level density in the critical
region (cf.~Fig.~\ref{fig-PeakPos}). Both checks work very well: numerically, we obtain exponents of $1.52$
and $0.73$, respectively.   

\section{Conclusions}
We have obtained numerical evidence, by extrapolating results of lattice
simulations to the continuum, that smeared Wilson loops in 4D continuum pure
$\SU(N)$ gauge theory undergo a large-$N$ phase transition at a critical loop
size. Furthermore, we have confirmed the expected universal properties of the
transition. For a complete presentation of our results (including data for
larger $N$) we refer to Refs.~\cite{LNprep, Lohmayer2011}.

\acknowledgments
RL and HN acknowledge partial support by the DOE under grant
number DE-FG02-01ER41165. We are grateful to R.~Narayanan who was
involved in the early stages of this project. 

\bibliographystyle{JHEP}
\bibliography{lohmayerLat2011}

\providecommand{\href}[2]{#2}\begingroup\raggedright\begin{thebibliography}{1}

\bibitem{Durhuus:1980nb}
B.~Durhuus and P.~Olesen, {\it {T}he spectral density for two-dimensional
  continuum {QCD}},  {\em Nucl. Phys.} {\bf B184} (1981) 461.

\bibitem{Narayanan:2006rf}
R.~Narayanan and H.~Neuberger, {\it {I}nfinite {N} phase transitions in
  continuum {W}ilson loop operators},  {\em JHEP} {\bf 03} (2006) 064,
  [\href{http://xxx.lanl.gov/abs/hep-th/0601210}{{\tt hep-th/0601210}}].

\bibitem{Narayanan:2007dv}
R.~Narayanan and H.~Neuberger, {\it {U}niversality of large {N} phase
  transitions in {W}ilson loop operators in two and three dimensions},  {\em
  JHEP} {\bf 12} (2007) 066, [\href{http://xxx.lanl.gov/abs/0711.4551}{{\tt
  arXiv:0711.4551}}].

\bibitem{Lohmayer:2009aw}
R.~Lohmayer, H.~Neuberger, and T.~Wettig, {\it {E}igenvalue density of {W}ilson
  loops in 2{D} {SU}({N}) {YM}},  {\em JHEP} {\bf 05} (2009) 107,
  [\href{http://xxx.lanl.gov/abs/0904.4116}{{\tt arXiv:0904.4116}}].

\bibitem{Neuberger:2008mk}
H.~Neuberger, {\it {B}urgers' equation in 2{D} {SU}({N}) {YM}},  {\em Phys.
  Lett.} {\bf B666} (2008) 106--109,
  [\href{http://xxx.lanl.gov/abs/0806.0149}{{\tt arXiv:0806.0149}}].

\bibitem{Lohmayer2011a}
R.~Lohmayer and H.~Neuberger, {\it {C}ontinuous smearing of {W}ilson {L}oops},
  {\em PoS} {\bf LAT2011} (2011) [\href{http://xxx.lanl.gov/abs/1110.3522}{{\tt
  arXiv:1110.3522}}].

\bibitem{LNprep}
R.~Lohmayer and H.~Neuberger {\em In preparation} (2011).

\bibitem{Lohmayer2011}
R.~Lohmayer and H.~Neuberger, {\it {N}on-analyticity in scale in the planar
  limit of {QCD}},  \href{http://xxx.lanl.gov/abs/1109.6683}{{\tt
  arXiv:1109.6683}}.

\end{thebibliography}\endgroup

\end{document}